\documentclass[prl,twocolumn,floats]{revtex4}
\usepackage{graphicx,epsfig}% Include figure files

\begin{document}
%\draft
%\preprint{}

\title{Electron correlation in the second Landau level; a competition between many, nearly degenerate quantum phases}

\author{J.S. Xia$^{1,2}$, W. Pan$^{3,2}$, C.L. Vincente$^{1,2}$, E.D. Adams$^{1,2}$, N.S. Sullivan$^{1,2}$,\\
H.L. Stormer$^{4,5}$, D.C. Tsui$^3$, L.N. Pfeiffer$^5$, K.W. Baldwin$^5$, and K.W. West$^5$}

\affiliation{$^1$University of Florida, Gainesville, Florida 32611}
\affiliation{$^2$National High Magnetic Field Laboratory, Tallahassee, Florida 32310}
\affiliation{$^3$Princeton University, Princeton, New Jersey 08544}
\affiliation{$^4$Columbia University, New York, New York 10027}
\affiliation{$^5$Bell Labs, Lucent Technologies, Murray Hill, New Jersey 07954}

\vskip5pc

\begin{abstract}

At a very low temperature of 9mK, electrons in the 2nd Landau level of an extremely high mobility two-dimensional electron
system exhibit a very complex electronic behavior. With varying filling factor, quantum liquids of different origins compete
with several insulating phases leading to an irregular pattern in the transport parameters. We observe a fully developed
$\nu=2+2/5$ state separated from the even-denominator $\nu=2+1/2$ state by an insulating phase and a $\nu=2+2/7$ and
$\nu=2+1/5$ state surrounded by such phases. A developing plateau at $\nu=2+3/8$ points to the existence of other
even-denominator states.

\end{abstract}


\date{\today}
\maketitle

Low-temperature electron correlation in the lowest Landau level (LL) of a two-dimensional electron system (2DES) separates
largely into two regions. At very low filling factor, $\nu \leq 1/5$, an insulating phase exists, which has now quite
convincingly been determined to be a pinned electron solid \cite{jiang90,ye02,review}. At higher filling factor $1 > \nu >
\sim 1/5$ the multiple sequences of fractional quantum Hall effect (FQHE) liquids \cite{tsui82,laughlin83,qhebook1,qhebook2}
dominate, which show the characteristic vanishing magneto resistance, $R_{xx}$, and quantized Hall resistance, $R_{xy}$, at
many odd-denominator rational fractional fillings $\nu=p/q$ \cite{note}. Altogether about fifty such FQHE states have been
observed in this region. Their multiple sequences can largely be described within the composite fermion (CF) model
\cite{jain89,hlr93,cfbook1,cfbook2}, with the exact origin of some higher order states still being argued. The electrical
behavior between FQHE states carries no particularly strong transport signature, being thought of as arising largely from the
conduction of excited quasiparticles of the neighboring FQHE states, with CF liquids occurring at some even-denominator
fractions.

At high LL's a very different pattern seems to emerge. There charge density wave (CDW) or liquid crystal like states
dominate, often referred to as electronic stripe and bubble phases \cite{lilly99a,du99,fogler02}. Characteristically these
states are pinned to the lattice, immobilizing the electrons of this LL, which leads to transport properties identical to
those of the neighboring integer quantum Hall effect (IQHE) states. FQHE states are absent in these high LL's, except for the
recent observation of two FQHE features in the third LL, at elevated temperatures \cite{gervais04}. Of course, high LL
fillings typically occur at lower magnetic fields and hence at poorer resolution of potential FQHE features. However, very
general theoretical arguments \cite{fukuyama79,fogler96} based on an increasing extent of the wavefunction with increasing LL
index, hence the increasing importance of exchange and the diminishing applicability of point-like interactions, clearly
support this trend.

It is in the 2nd LL where electron liquids and electron solids collide. The larger extent of the wavefunction as compared to
the lowest LL and its additional zero allows for a much broader range of electron correlations to be favorable, leading to an
ever changing competition between multiple electronic phases as the filling factor is varied and as the temperature is
lowered.

An early example of the variety of electron correlations encountered in the 2nd LL is the even-denominator $\nu=2+1/2$ FQHE
state \cite{willett87,pan99a}. From numerical calculations \cite{morf98,rezayi00}, supported by several experimental facts
\cite{pan99b,lilly99b}, there is good evidence that this $\nu=5/2$ state consists of pairs of composite fermion
\cite{moore91,greiter91,scarola00a}. As a function of tilt -- equivalent to a compression of the wavefunction perpendicular
to the 2D plane -- the FQHE phase gives way to a pinned, electrically anisotropic state as it appears in higher LL, thereby
revealing the close balance between FQHE liquids and pinned CDW states in the 2nd LL.

This delicate balance has been further highlighted through the observation of a reentrant integer quantum Hall effect (RIQHE)
in the 2nd LL \cite{eisenstein02}. At temperatures below 50mK the Hall trace between the $\nu=3$ and the $\nu=4$ IQHE plateau
does not follow the usual, almost classical Hall line. Instead, at four regions between $\nu=3$ and 4 it reverts to the level
of the integer Hall plateaus; two to its upper value of h/3e$^2$ and two to its lower value of h/4e$^2$. The equivalent
effect is seen in the lower spin state of the 2nd LL. These observations still lack a definitive explanation, however, an
interpretation in terms of a pinned solid (CDW or liquid crystal) phase appears inescapable
\cite{du,eisenstein02,shibata03,goerbig04}. These abrupt returns from a smooth Hall trace at higher temperatures to the IQHE
at very low temperatures underscores the fragility of these phases. In the ``windows'' between these pinned electronic
states, along the stretches of a steadily rising classical Hall line, so far only the $\nu=3+1/5$ and 4-1/5 FQHE
(equivalently $\nu=2+1/5$ and 3-1/5) appeared, apart from the prominent even-denominator states at half-fillings. These
observations indicate that the FQHE liquids in the 2nd LL have largely given way to insulating electronic many particle
states.

Our recent data, presented in this communication, point to a yet more complex state of affairs in the 2nd LL. At very low
temperatures of 9mK in a very high-mobility 2DES specimen, we observe several new FQHE liquid states competing with the RIQHE
states. In fact, while present at higher temperatures some FQHE states seem to be destroyed by the RIQHE at lower
temperatures. We observe well quantized FQHE states at $\nu=2+1/3$ and 2+2/3 in coexistence with the RIQHE and, quite
importantly, a new, clearly quantized FQHE state at $\nu=2+2/5$, the nature of such a state being theoretically intensely
debated. Finally, there is evidence for the presence of a second even-denominator FQHE at a filling of $\nu=2+3/8$ in our
data whose origin is also enigmatic.

In our experiments we used a 30~nm wide quantum well that is delta-doped on both sides of the well at a setback distance of
100nm. The electron density is $n=3 \times 10^{11}$ cm$^{-2}$ and the mobility is $\mu=31 \times 10^6$ cm$^2$/Vs. These
values are established after illumination of the specimen with a red light emitting diode at low temperatures. The ultra-low
temperature experiments were carried out in a demagnetization/dilution refrigerator combination described in Ref.
\cite{pan99a}. Standard low-frequency lock-in techniques ($\sim 7Hz$) were utilized to measure the magnetoresistance $R_{xx}$
and the Hall resistance $R_{xy}$, with an excitation current of 1nA, insuring that electron heating by the current was not a
limiting factor.

Fig.1 shows the diagonal resistance and the Hall resistance between filling factor $\nu=2$ and 3, taken at the base
temperature of $T$ = 9mK. At the extremes of the horizontal axis the standard quantized Hall plateaus in $R_{xy}$ and
vanishing resistance in $R_{xx}$ are just visible. Between these field values, $R_{xx}$ exhibits the typical spiky behavior
seen in low temperature measurements on 2DES. The Hall trace, however, is very unusual. Instead of moving monotonically and
in a stair-like fashion from h/3e$^2$ to h/2e$^2$, $R_{xy}$ along several stretches of filling factor returns to the value of
the neighboring IQHE plateaus. These are the features of the RIQHE. In between the RIQHE we observe now several plateaus with
fractional quantum number. (1) Fully developed FQHE states are observed at $\nu=2+1/2$, 2+1/3, and 2+2/3, all having wide
Hall plateaus quantized to better than 1 part in 10$^4$. (2) For the first time, a fully developed FQHE state at $\nu=2+2/5$
is observed, showing a quantized Hall plateau and vanishingly small $R_{xx}$. (3) A new even-denominator FQHE state seems to
be developing at $\nu=2+3/8$ as deduced from a deep minimum in $R_{xx}$ and a Hall plateau value within 0.2\% of $R_{xy}$ =
h/e$^2$/(2+3/8). (4) One of the RIQHE plateaus is split into two plateaus around $B \sim 5.7$~T, corresponding to filling
factor $\nu=2+2/7$, where simultaneously a minimum occurs in $R_{xx}$. These features signal a complex competition between
the FQHE state and the RIQHE state. (5) Finally, a strong $\nu=2+4/5$ FQHE state is seen at $B \sim 4.6$~T. At 9mK, there is
no FQHE at $\nu=2+1/5$. Rather $R_{xx}$ shows a peak. The rising Hall resistance at the B field just below $\nu=2+1/5$ seems
to indicate that another RIQHE is developing.

The fully developed $\nu=2+2/5$ FQHE state is observed for the first time. At our lowest temperature of 9mK the accuracy of
the $R_{xy}$ quantization is better than 0.02\% (using $R_{xy}$ at $\nu=5/2$ as a reference) and $R_{xx}$ reaches a low value
of $\sim$ 5 ohm (Fig. 2a). The formation of a true FQHE state is corroborated by the temperature dependence of d$R_{xy}$/dB
which reaches zero at $\sim$ 10~mK, as shown in Fig. 2c. The energy gap of this new state is $\sim$ 70~mK as determined from
the T-dependence of $R_{xx}$ in Fig. 2b.

The energy gap of the new $\nu=2+2/5$ state is disproportionately small. In the lowest LL the energy gap of the 2/5 state is
about half of the gap at $\nu=1/3$ \cite{boebinger87,du93}, whereas in this 2nd LL, the energy gap at 2/5 is only 1/10 of
that at $\nu=1/3$ ($\Delta_{7/3} \sim 0.6$~K). As to its origin, a first assumption would be the $\nu=2+2/5$ state to be a
hierarchical daughter state of the $\nu=2+1/3$ state, as in the lowest LL \cite{haldane83,halperin84}. However, Read and
Rezayi \cite{read99} found, for coulombic interactions, vanishing overlap of such a hierarchical state with the exact ground
state. This result is corroborated in a numerical study by Morf and d'Ambrumenil \cite{morf95}, which determined that no
hierarchical state was stable between $\nu=2+2/3$ and 2+1/3. Also within the composite fermion model the traditional
$\nu=2+2/5$ state is unstable as seen from the collapse of its neutral excitation in numerics \cite{scarola00b}. Taken
together, these studies strongly suggest that the $\nu=2+2/5$ state is not a conventional FQHE state.

Several non-conventional ground state wavefunctions have been proposed for FQHE states in the 2nd LL. Among them, the
parafermioinc state of FQHE is the most exciting \cite{read99}. In this model, a number $k$ ($\leq$ 2) of electrons form
clusters and it is the condensation of these clusters that gives rise to a FQHE state at $\nu=2+k/(k+2)$. This model would
explain the existence of the already observed FQHE states at $\nu=2+2/3$ ($k=4$) and $\nu=2+1/3$, by particle-hole symmetry.
It also predicts the next new FQHE state to occur at $\nu=2+3/5$ and its particle hole conjugate state at $\nu=2+2/5$.
Indeed, we observe the $\nu=2+2/5$ FQHE state in our experiment. However, there is no evidence of a FQHE state at
$\nu=2+3/5$. It is not clear whether this absence is due to a broken particle-hole symmetry, is related to the latter state
residing at lower $B$-field and hence being weaker, or is due to an asymmetry in the nearby RIQHE. Consistent with the
parafermionic model, no other FQHE states occur between $\nu=2+2/5$ and 2+1/2. However, their absence could also be a result
of the RIQHE being the true ground state in this regime.

In fact, the formation of k-electron clusters provides a compelling scenario to interpret the data in this filling factor
range. Although the origin of RIQHE remains unclear, it is believed to arise from collective freezing of electron clusters
\cite{eisenstein02,du,shibata03,goerbig04}. Then the crossover from the $\nu=2+2/5$ state to the RIQHE state may represent a
quantum phase transition from an electron-cluster liquid to an electron-cluster solid, resembling the phase transition from
electron liquid to electron solid in the lowest LL \cite{jiang90,ye02,review}. Numerical studies predict such a transition at
$\nu \sim 2+0.37$ \cite{shibata03,goerbig04}, which would be to the right of $\nu=2+2/5$ in Fig.1. Instead it occurs just to
the left. This difference could well result from the finite thickness of 2DES's in real samples, which generally stabilizes
the liquid states in the 2nd LL \cite{belkhir95}, but is not included in the calculations.

Beyond the $\nu=2+2/5$ FQHE state we clearly observe a new, developing even-denominator FQHE state at $\nu=2+3/8$. As shown
in Fig.2a, together with a deep minimum in $R_{xx}$, a Hall plateau is forming. It is centered at $R_{xy}$ = h/e$^2$/(2+3/8)
within 0.2\%. Moreover, d$R_{xy}$/d$B$ at $\nu=2+3/8$ decreases with decreasing temperature (Fig.2c). Extrapolating its
temperature dependence, a flat Hall plateau is expected to form at an electron temperature of 2-3 mK.

At the present time, the physical origin of the $\nu=2+3/8$ state remains unclear. The possibility of a FQHE state at
$\nu=3/8$ in the lowest LL was considered recently \cite{scarola02,quinn03,lopez04}. It will be interesting to see whether a
similar mechanism could also be responsible for a $\nu=2+3/8$ fraction in the 2nd LL. On the other hand, the existence of the
$\nu=2+3/8$ state is not consistent with the parafermionic model, which would predict a FQHE at $\nu=2+3/4$ ($k=6$). However,
the Hall resistance of Fig.1 shows no such feature. Detailed numerical calculations will be required to resolve the origin of
the $\nu=2+3/8$ state.

In addition, the $\nu=2+2/7$ and $\nu=2+1/5$ states provide more striking evidence for the competition between solid and
liquid phase in the second LL, see Fig. 3. At temperatures above $\sim$ 40~mK, the $\nu=2+1/5$ state is well developed in
$R_{xx}$ and in $R_{xy}$. Yet on lowering $T$, the neighboring $\nu=2$ IQHE is taking over, transforming the $R_{xy}$ plateau
into a local minimum, which lifts off from the quantized value and also shifts to higher n. At the lowest $T=9$~mK of Fig. 1,
the remaining $\nu=2+1/5$ features are clearly surrounded by the RIQHE whose Hall value tends toward h/2e$^2$. This behavior
is yet more prominent around $\nu=2+2/7$, where in Fig. 3 at 16~mK the quantized Hall value is approached in a very narrow
downward spike. However, the state is increasingly destroyed on lowering T as is evident from the $R_{xy}$ trace in Fig. 1.
Furthermore, the $T$ dependences of the RIQHE on both sides of $\nu=2+2/7$ (as well as $\nu=2+1/5$) are rather different,
possibly indicating two different solid states on either side of the FQHE liquid. Overall, the observed features resemble the
behavior at very low filling factor in the lowest LL \cite{pan02}. There, the observed melting transition was attributed to
the decrease of the free energy of the liquid as compared to the solid due to the increasing population of states in the
roton minimum as T is raised. A similar mechanism may be responsible for the observed crossover in the 2nd LL.

In summary, at the sample temperature of 9~mK, we observe a very complex electronic transport behavior in the 2nd LL of a
high quality 2DES, pointing to a close balance between liquid and solid ground state energies. A well quantized $\nu=2+2/5$
state is observed in spite of its absence in numerical calculations based on traditional e-e correlation. This may be
evidence for the existence of parafermionic behavior. Some aspects of the competition between solid and liquid states are
reminiscent of features at very low n in the lowest LL and may have similar explanations. All together, e-e correlation in
the 2nd LL proves to be rather distinct from the lowest LL with many competing quantum phases closely spaced in energy.

We thank Rui Rui Du, Duncan Haldane, Mark Goerbig, Jainendra Jain, Ed Rezayi, and Steve Simon for useful discussions. The
work at Princeton was supported by the AFOSR, the DOE, and the NSF. The work at Columbia was supported by DOE, NSF and by the
W.M. Keck Foundation. Experiment was carried out at the high B/T facilities of the National High Magnetic Field Laboratory,
which is supported by NSF Cooperative Agreement No. DMR-0084173 and by the State of Florida.

\begin{figure} [h]
\centerline{\epsfig{file=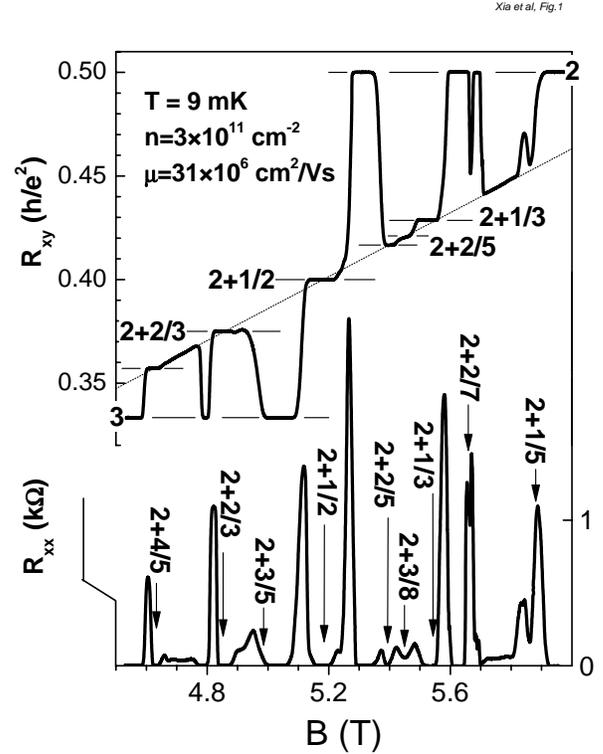,width=8.5cm}}
\caption{
$R_{xx}$ and $R_{xy}$ between $\nu=2$ and $\nu=3$ at 9mK. Major FQHE states are marked by arrows. The horizontal
lines show the expected Hall value of each FQHE state. The dotted line is the calculated classical Hall resistance.
}
\end{figure}

\begin{figure} [h]
\centerline{\epsfig{file=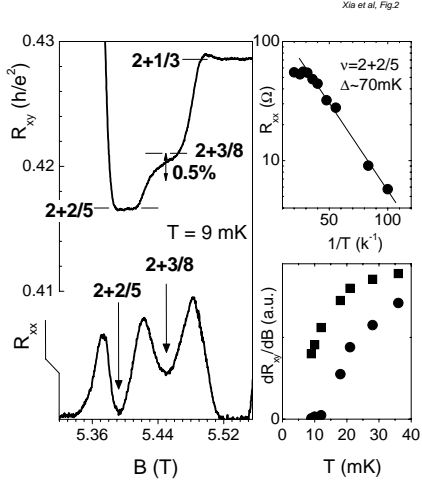,width=6cm}}
\caption{
(a) $R_{xx}$ and $R_{xy}$ around $\nu=2+2/5$ and $\nu=2+3/8$. The horizontal lines indicate expected values of the
Hall plateaus of these two states. The vertical arrow shows the 0.5\% deviation from the expected value for the $\nu=2+2/8$
state. (b) Arrhenius plot for the $R_{xx}$ minimum at $\nu=2+2/5$. The line is a linear fit. (c) d$R_{xx}$/d$B$ vs. T for the
$\nu=2+2/5$ (solid circles) and $\nu=2+3/8$ (solid squares) states.
}
\end{figure}

\begin{figure} [h]
\centerline{\epsfig{file=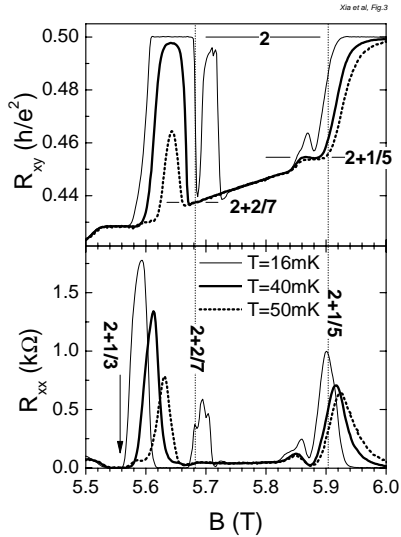,width=6cm}}
\caption{
Temperature dependence of $R_{xx}$ and $R_{xy}$ around the split Hall plateau and $\nu=2+1/5$. The vertical lines
mark the B field positions of the $\nu=2+1/5$ and 2+2/7 states. The horizontal lines mark the expected Hall resistance values
for $\nu=2+2/7$, 2+1/5, and 2.
}
\end{figure}


\begin{thebibliography} {90}
\bibitem{note}
An electron-hole symmetric solid of holes is expected for $\nu$ close to $\nu=1$.
\end{thebibliography}
\end{document}